\documentclass[journal,twoside,web]{ieeecolor}
\usepackage{tmi2}
\usepackage{cite}
\usepackage{amsmath,amssymb,amsfonts}
\usepackage{algpseudocode}
\usepackage[ruled,vlined]{algorithm2e}
\usepackage{graphicx}
\usepackage{textcomp}
\usepackage{float}
\usepackage{adjustbox}
\usepackage{mathtools}
\SetKwInput{KwInput}{Input}                
\SetKwInput{KwOutput}{Output}              

\usepackage[font=footnotesize,justification=justified,belowskip=2pt,aboveskip=2pt]{caption}
\usepackage[math]{cellspace}
\usepackage[utf8]{inputenc}
\usepackage{url}
\usepackage{multirow}
\usepackage{xcolor}
\usepackage{comment}
\usepackage{hyperref}
\hypersetup{colorlinks=true,citecolor=teal,linkcolor=blue,urlcolor=blue}

\makeatletter
\def\widebreve{\mathpalette\wide@breve}
\def\wide@breve#1#2{\sbox\z@{$#1#2$}%
     \mathop{\vbox{\m@th\ialign{##\crcr
\kern0.08em\brevefill#1{0.8\wd\z@}\crcr\noalign{\nointerlineskip}%
                    $\hss#1#2\hss$\crcr}}}\limits}
\def\brevefill#1#2{$\m@th\sbox\tw@{$#1($}%
  \hss\resizebox{#2}{\wd\tw@}{\rotatebox[origin=c]{90}{\upshape(}}\hss$}

\newcommand*{\bigCup}{\mathop{\mathpalette\big@Cup\relax}\slimits@}
\newcommand*{\big@Cup}[2]{%
   \setbox\z@=\hbox{\m@th$#1\Cup$}%
   \setbox\z@=\vtop{\vbox{\kern.2\ht\z@\copy\z@}\kern.1\ht\z@}%
   \setbox\tw@=\hbox{\m@th$#1\bigcup$}%
   \vcenter{\hbox{\resizebox{!}{1.4\ht\tw@}{\box\z@}}}%
}
\makeatletter

\def\SP#1{\textsuperscript{#1}}
\def\SB#1{\textsubscript{#1}}
\def\BibTeX{{\rm B\kern-.05em{\sc i\kern-.025em b}\kern-.08em
    T\kern-.1667em\lower.7ex\hbox{E}\kern-.125emX}}

\begin{document}

\title{Learning Fourier-Constrained Diffusion Bridges for MRI Reconstruction}
\author{M. Usama Mirza, Onat Dalmaz, Hasan A. Bedel, Gokberk Elmas, Yilmaz Korkmaz, Alper Gungor,\\ Salman UH Dar, and Tolga \c{C}ukur$^*$
\vspace{-1.2cm}
\\
\thanks{This study was supported in part by TUBA GEBIP 2015 and BAGEP 2017 fellowships, and by a TUBITAK 1001 Grant 121E488 awarded to T. \c{C}ukur  (Corresponding author: Tolga \c{C}ukur, cukur@ee.bilkent.edu.tr).}
\thanks{Authors are with the Dept. of Electrical-Electronics Engineering and National Magnetic Resonance Research Center (UMRAM), Bilkent University, Ankara, Turkey, 06800. T. \c{C}ukur is also with the Neuroscience Graduate Program Bilkent University, Ankara, Turkey, 06800. }
}

\maketitle
\begin{abstract}
    Deep generative models have gained recent traction in accelerated MRI reconstruction. Diffusion priors are particularly promising given their representational fidelity. Instead of the target transformation from undersampled to fully-sampled data required for MRI reconstruction, common diffusion priors are trained to learn a task-agnostic transformation from an asymptotic start-point of Gaussian noise onto the finite end-point of fully-sampled data. During inference, data-consistency projections are injected in between reverse diffusion steps to reach a compromise solution within the span of both the trained diffusion prior and the imaging operator for an accelerated MRI acquisition. Unfortunately, performance losses can occur due to the discrepancy between target and learned transformations given the asymptotic normality assumption in diffusion priors. To address this discrepancy, here we introduce a novel Fourier-constrained diffusion bridge (FDB) for MRI reconstruction that transforms between a finite start-point of moderately undersampled data and an end-point of fully-sampled data. We derive the theoretical formulation of FDB as a generalized diffusion process based on a stochastic degradation operator that performs random spatial-frequency removal. We propose an enhanced  sampling algorithm with a learned correction term for soft dealiasing across reverse diffusion steps. Demonstrations on brain MRI indicate that FDB outperforms state-of-the-art methods including non-diffusion and diffusion priors. 
\end{abstract}
\vspace{-0.1cm}
\begin{IEEEkeywords} diffusion, bridge, generative, deep learning, MRI, reconstruction \vspace{-0.2cm}
\end{IEEEkeywords}

\bstctlcite{IEEEexample:BSTcontrol}

\vspace{-0.1cm}
\section{Introduction}
MRI is a diagnostic powerhouse with exceptional soft-tissue contrast that suffers from long scan times. Acceleration via undersampled acquisitions helps lower operational costs and susceptibility to patient motion, albeit an ill-posed inverse problem must be solved to reconstruct images \cite{lustig2007sparse}. Given their high sensitivity, image priors based on deep learning have become pervasive in MRI reconstruction over the years \cite{Dong2020spm,sandino2020SPM,hammernik2022physicsdriven,Lam2023}. A prevalent framework uses task-specific priors that capture a dealiasing transformation from undersampled to fully-sampled data \cite{Wang2016,Schlemper2017,Hammernik2017,Zhu2018,MoDl}. The transformation is expressed as a conditional model often trained for a specific imaging operator (i.e., with fixed acceleration rate and sampling density) \cite{Dar2017,Kwon2017,Mardani2019b,ramzi2023unrolled,wang2022bjork}. Although task-specific priors typically offer high within-domain performance, their reliability can be compromised under significant domain shifts between training and test sets (e.g., due to shifts in sampling density or dataset) \cite{liu2019mrm,wang2023xiaobo,pramanik2023adamodl}.

An alternative framework to improve reliability uses task-agnostic priors based on generative models \cite{tezcan2018mr}, trained to capture the marginal distribution of fully-sampled data divorced from the imaging operator \cite{narnhofer2019inverse,korkmaz2022unsupervised}. Of particular promise, diffusion priors use an asymptotic process to transform between Gaussian noise as the asymptotic start-point and the data as the end-point \cite{jalaln2021nips,chung2022score}. For reconstruction, sampling is initiated on a random noise image, and alternating projections through the prior and the imaging operator can be performed to reach a compromise solution \cite{luo2022uncertainty,xie2022kspace,peng2022towards,cao2022highfrequency,gungor2022adaptive,peng2023oneshot}. Note that the normality assumption elicits divergence between the transformations implemented by the prior and the imaging operator, potentially lowering optimization efficiency \cite{chung2022cvpr,cui2023spiritdiffusion,ozturkler2023regularization}. While cold diffusion priors that use asymptotic processes based on known, deterministic image degradations have been proposed \cite{bansal2022cold,daras2022soft,lee2022progressive}, they have been suggested to offer similar image quality with common diffusion priors in MRI reconstruction \cite{huang2023cdiffmr}. These restrictions can hinder diffusion-based MRI reconstruction from  realizing its full potential.  

An emerging approach to boost flexibility in generative modeling is diffusion bridges that perform a delimited transformation between two arbitrary distributions by relinquishing the normality assumption in common diffusion priors \cite{liu2023i2sb}. Instead of Gaussian noise, recent diffusion bridges devised for inverse problems have taken the finite start-point as a Dirac-delta distribution pertaining to images degraded with the known, deterministic imaging operator  \cite{fabian2023diracdiffusion,delbracio2023inversion,chung2023direct}. Intermediate degradations in forward diffusion steps have been assumed as either unconstrained (i.e., learnable) or constrained to a convex combination of start- and end-points \cite{delbracio2023inversion,chung2023direct}. While promising results have been reported in computer vision tasks \cite{chung2023direct}, to our knowledge, diffusion bridges remain relatively unexplored in MRI reconstruction. This might be non-trivial given the random undersampling patterns pervasively used in accelerated MRI that elicit highly stochastic image degradations \cite{lustig2007sparse}.  

Here, we introduce a novel Fourier-constrained diffusion bridge (FDB) for improved performance in MRI reconstruction (Fig. \ref{fig:ddpm_vs_fdb}). FDB leverages a generalized diffusion process to map between fully-sampled and undersampled data. In the forward direction, a stochastic degradation operator removes a random set of spatial frequencies per step, scheduled to follow a peripheral-to-central k-space order from the end- to the start-point. Unlike common or cold diffusion priors with an asymptotic start-point, FDB uses a finite start-point based on a moderate level of degradation (i.e., undersampling). Unlike previous diffusion bridges based on deterministic degradation, FDB employs stochastic degradations. For inference, starting with the least-squares reconstruction of an undersampled test acquisition, an enhanced sampling algorithm is devised for FDB with a learned correction term to achieve soft dealiasing. Comprehensive demonstrations on brain MRI datasets indicate the superior performance of FDB against previous non-diffusion and diffusion priors. Code for FDB is available at {\small \url{https://github.com/icon-lab/FDB}}. 

\vspace{0.2cm}
\subsubsection*{\textbf{Contributions}}
\begin{itemize}
    \item To our knowledge, FDB is the first diffusion bridge for accelerated MRI reconstruction in the literature.
    \item FDB leverages a novel generalized diffusion process based on random spatial-frequency removal to map between fully-sampled and undersampled data. 
    \item A sampling algorithm with a learned correction term is introduced for FDB that enables soft dealiasing across diffusion steps for enhanced image recovery. 
\end{itemize}

\vspace{-0.05cm}
\section{Related Work}
\subsection{MRI Reconstruction}
Deep learning frameworks have empowered leaps in image quality over traditional methods for accelerated MRI reconstruction. A mainstream approach relies on conditional models that are trained to map undersampled acquisitions as input onto fully-sampled acquisitions as output \cite{Hyun2018,Quan2018c,Yu2018c,KikiNet}. The resultant task-specific priors capture a dealiasing transformation particular to the prescribed imaging operator \cite{Dar2017,Yoon2018}. Despite their prominence, many task-specific priors produce a deterministic output limiting their representational capacity \cite{Liu2020mrm,dar2020prior,guo2022reconformer}, and they might be rendered ineffective under domain shifts in imaging operators due to on-the-fly changes during scan sessions to meet clinical demand \cite{tezcan2018mr,yaman2021zero}.

To improve generalization, an alternative approach employs task-agnostic priors based on generative models that transform random latent variables onto high-quality MR images decoupled from the imaging operator \cite{narnhofer2019inverse,pixelcnnrecon,darestani2021accelerated}. To reconstruct undersampled data, the image sets spanned by the prior and the imaging operator can be consolidated via alternating projections \cite{slavkova2023untrained,elmas2022federated}. This consolidation naturally rests on the representational capacity of the prior \cite{gungor2022adaptive}. Diffusion priors have recently been adopted given their superior representation, enabled via a stochastic process that gradually transforms between Gaussian noise and fully-sampled data \cite{ho2020denoising,jalaln2021nips,chung2022score,song2022}. While promising results have been reported in MRI reconstruction \cite{luo2022uncertainty,xie2022kspace,peng2022towards}, common diffusion priors rely on an asymptotic normality assumption to form the start-point of the diffusion process \cite{ho2020denoising}. This causes divergence between the transformation captured by the prior and the target transformation from undersampled to fully-sampled data, potentially compromising efficiency and performance \cite{cui2023spiritdiffusion,ozturkler2023regularization}. 

To address this limitation, here we introduce the first diffusion bridge for MRI reconstruction in the literature to our knowledge. The proposed FDB prior leverages a stochastic degradation operator that randomly discards spatial-frequency components from fully-sampled data, starting in peripheral k-space and gradually progressing towards central k-space. Unlike common diffusion priors with Gaussian noise as an asymptotic start-point, FDB employs a finite start-point based on moderately undersampled data. This generalized diffusion process enables FDB to capture task-relevant information in its diffusion prior for performant MRI reconstruction.

\begin{figure*}[!t]
       \vspace{-0.25cm}
        \captionsetup{justification=justified, singlelinecheck=false}
        \ifx\fastCompile\undefined
        \begin{minipage}{0.25\textwidth}
        \caption{\textbf{a)} In forward steps, common diffusion priors add random noise onto a clean image ($x_0$) to reach an asymptotic start-point of isotropic Gaussian noise ($x_T$). For reconstruction, starting with a random $x_T$, reverse steps with the diffusion prior for denoising are interleaved with data-consistency projections. \textbf{b)} In contrast, FDB leverages a stochastic frequency-removal operator to map Fourier-domain data for a fully-sampled acquisition ($X_0$) onto a finite start-point ($X_{Tf}$) corresponding to $R'$-fold undersampled data. To reconstruct an $R$-fold undersampled acquisition $y$, sampling is initiated with the least-squares solution ${x}_{T_r} = A^H y$, where $A^H$ denotes the Hermitian adjoint of the imaging operator $A$, and $T_r = \lfloor T_f \frac{(R-1)R'}{(R'-1)R} \rfloor$. Reverse steps with FDB for frequency imputations are interleaved with data-consistency projections.}  
        \label{fig:ddpm_vs_fdb}
        \end{minipage}
        \hspace{0.05\linewidth}
        \begin{minipage}{0.75\textwidth}
        \includegraphics[width=0.835\textwidth]{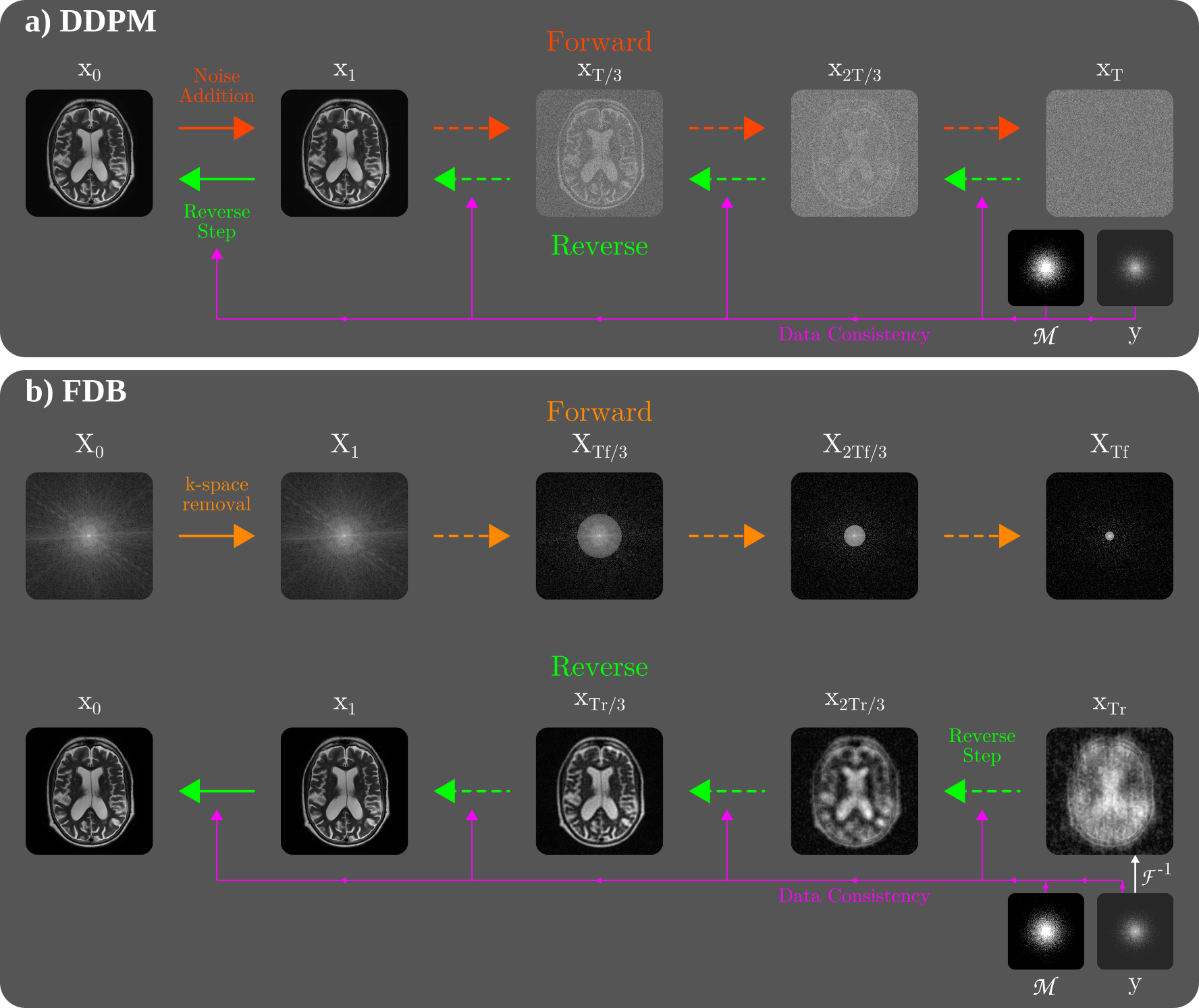}
        \end{minipage}
\end{figure*}

\subsection{Diffusion Bridges}
Improvements to common diffusion priors have recently been sought to increase flexibility in generative modeling. Computer vision studies have proposed cold diffusion priors built on deterministic degradation operators such as blurring or downsampling in lieu of noise addition \cite{bansal2022cold,daras2022soft,lee2022progressive}. To improve sampling efficiency, a recent study on single-coil MRI reconstruction has independently considered a cold diffusion prior based on k-space undersampling \cite{huang2023cdiffmr}. Yet, cold diffusion methods often rely on deterministic degradations assumed to remain fixed between training-test sets, which reduces the prior's stochasticity and might limit generalization \cite{bansal2022cold}. They also employ asymptotic start-points at high degradation levels \cite{huang2023cdiffmr}, which can limit performance by elevating difficulty in reverse diffusion sampling \cite{balaji2022ediffi}. In contrast, FDB employs a finite, stochastic start-point of moderately undersampled data obtained via random frequency removal. 

An emerging alternative for inverse-problem solutions is diffusion bridges that directly transform between degraded and clean images. Primarily devised for known, deterministic image degradations, existing diffusion bridges define the start-point as a Dirac-delta distribution reflecting degraded images with limited stochasticity \cite{fabian2023diracdiffusion,delbracio2023inversion}. To improve representational diversity, FDB instead uses a stochastic start-point by removing a random set of frequencies during forward diffusion, and randomly varying the removed set across training samples. The definition of forward diffusion mappings is a unique technical aspect of FDB. While unconstrained bridges with learnable forward mappings can suffer performance losses due to elevated model complexity \cite{liu2023i2sb}, recent studies have considered constrained bridges with intermediate samples taken as a convex combination of images at start- and end-points \cite{fabian2023diracdiffusion,delbracio2023inversion,chung2023direct}. Yet, this can elicit a linear averaging of degraded and clean images, which diverges from the nature of accelerated MRI scans where a binary selection is exercised on acquired k-space points. Instead, FDB leverages a task-oriented constraint by removing a monotonically growing set of frequencies across forward steps based on binary selection.

\section{Theory}
\subsection{MRI Reconstruction with Common Diffusion Priors}

Reconstruction aims to recover a high-quality MR image $x$ of an anatomy given undersampled k-space acquisition $y$:
\begin{equation}
        \label{eq:undersampling}
        A x + \varepsilon = y ,
\end{equation}
where $A$ is the imaging operator, $\varepsilon$ is measurement noise. $Ax$=$M\mathcal{F}(Sx)$ comprises the effects of the sampling pattern $M$, Fourier encoding $\mathcal{F}$, and coil sensitivities $S$. Reconstruction is an ill-posed inverse problem, so a regularization term is typically employed to obtain a faithful solution $\hat{x}$ \cite{lustig2007sparse}:
\begin{equation}
        \label{eq:reconstruction}
        \hat{x} = \underset{x}{\operatorname{min}} \left\| A x - y \right\|^2 + G(x,y),
\end{equation}
where $G(x,y)$ is the regularizer that incorporates prior information on the distribution of MR images. Task-specific priors are typically expressed as a conditional regularizer $G(x|y)$ that captures a dealiasing transformation. Task-agnostic priors are instead expressed as a marginal regularizer $G(x)$ independent of the imaging operator to improve generalization. 

Recently adopted as task-agnostic priors for MRI \cite{jalaln2021nips,chung2022score}, common diffusion priors degrade a clean image $x_0$ over $T$ steps in the forward direction to obtain an isotropic noise sample $x_T$. The sample $x_{t}$ at step $t \in (0 \mbox{ }T]$ is \cite{ho2020denoising}:
\begin{eqnarray}
    \label{eq:diff_forward}
    x_{t}=\sqrt{1-\beta_{t}} x_{t-1}+ \sqrt{\beta_t} z,
\end{eqnarray}
where $\beta_t$ is noise variance, $z\sim\mathcal{N}(0,\mathrm{I})$ follows a Gaussian distribution, $\mathrm{I}$ is an identity matrix. In the reverse direction, image samples are generated via repeated denoising. One approach uses a recovery operator $G_{\theta}$ to predict $\Tilde{x}_0 = G_{\theta}(\hat{x}_t, t)$, and draw a denoised sample $\dot{x}_{t-1}$ from the normal distribution $q(x_{t-1} | x_{t},\Tilde{x}_0)$ \cite{nichol2021improved}. Training objective for $G_{\theta}$ is given as:
\begin{eqnarray}
\label{eq:common_obj}
   L_{common} = \mathbb{E}_{t, x_{(0,t)}} [\| G_{\theta}(x_t,t) - x_0 \|^2],
\end{eqnarray}
where $\mathbb{E}$ is expectation, $t \sim U[1,T]$ follows a uniform distribution, $x_0, x_t \sim q(x_0,x_t)$ are generated via the forward process. For MRI reconstruction, starting with a random noise sample $\hat{x}_T \sim \mathcal{N}(0,\mathrm{I})$, reverse diffusion steps \cite{nichol2021improved} can be interleaved with data-fidelity projections \cite{peng2022towards}:
\begin{align}
        \label{eq:reverse_step}
        & \dot{x}_{t-1} = \frac{\sqrt{\gamma_t}(1-\bar{\gamma}_{t-1})}{1-\bar{\gamma}_{t}} \hat{x}_t + 
        \frac{\beta_t \sqrt{\bar{\gamma}_{t-1}}}{1-\bar{\gamma}_{t}} \Tilde{x}_0 + \sqrt{\frac{1-\bar{\gamma}_{t-1}}{1-\bar{\gamma}_{t}}\beta_t} z,\\[-0.1cm]
        \label{eq:dataconsistency}
        & \hat{x}_{t-1} = \dot{x}_{t-1} + A^H (y - A \dot{x}_{t-1}),
\end{align}
where $\gamma_t=1 - \beta_t$, $\bar{\gamma}_t = \prod_{\tau=[0,1,..,t]} \gamma_t$, and $A^H$ is the Hermitian adjoint of the imaging operator $A$.

\subsection{Fourier-Constrained Diffusion Bridges}
Instead of an asymptotic start-point of Gaussian noise, the proposed FDB prior uses moderately undersampled data as its finite start-point. To reach the start-point, FDB introduces a stochastic degradation operator in Fourier domain that removes a random set of frequency components in forward steps (Fig. \ref{fig:ddpm_vs_fdb}). We derive the training objective of the recovery operator based on the interpretation that it forms a generalized diffusion process in image domain \cite{daras2022soft}. We then present an enhanced sampling algorithm incorporating a temporal-correction term to mediate soft dealiasing during reverse steps.
\vspace{-0.2cm}
\subsubsection{Diffusion process}
In Fourier domain, FDB maps between an end-point $X_0=\mathcal{F} (S^H \mathcal{F}^{-1} (y_{full}))$ taken as coil-combined Fourier-domain data derived from fully-sampled acquisitions, and a start-point $X_{T_f}$ at time step $T_f$ taken as coil-combined data degraded at level $R'$ via frequency removal ($R'$ is analogous to MRI acceleration rate). 

\textbf{Degradation operator.} In a forward step, the degradation operator is cast as a diagonal frequency-removal matrix $\mathbf{\Lambda}_t$. Given a two-dimensional Cartesian k-space grid $K$ with $N_K$ frequency components, let $k \in [1 \mbox{ } N_K]$ denote linear component index, and $r(k) \in [0 \mbox{ } r_{\mathrm{max}}]$, $\phi(k) \in [0 \mbox{ } 2\pi)$ denote the k-space radius and angle of the $k$th component. At step $t$, $\left.\mathbf{\Lambda}_t(k,k)\right\vert_{k\in S_t} = 0$ for $n$ components in the randomly selected set $S_t$, and $\left. \mathbf{\Lambda}_t(k,k)\right\vert_{k\notin S_t} = 1$. To improve the prior via monotonically increasing degradation across time \cite{fabian2023diracdiffusion}, we observe that components selected at $t$ should not overlap with previously selected components at $\{t-1,..,1\}$. To respect the energy distribution of MRI data, we observe that the selection should follow a peripheral-to-central k-space order as typical in accelerated MRI \cite{lustig2007sparse}. We enforce these selection guidelines via a spatio-temporal point process over k-space and $t$: 
\begin{align}
      \label{eq:k-selection}
        S_t &= \Big\{ (k_1,..,k_n) : \enspace k_i \sim U [1,N_K] \enspace \mbox{for } i=(1,..,n), \nonumber \\[-0.1cm]
        & \qquad \qquad \qquad \quad \enspace k_i \notin {\bigcup}_{\tau=1}^{t-1} S_{\tau}, \enspace r(k_i) > \bar{r}_t  \Big\},
\end{align}
where $U$ is uniform distribution, $r_{max}$ is the maximum radius in $K$, and $\bar{r}_t$ is a radius threshold scheduled as: 
    \begin{equation}
        \label{eq:radius}
        \bar{r}_t = r_{\mathrm{max}} - r_{\mathrm{max}} ( 1 - \sqrt{R'})(t/{T_f}).
    \end{equation}
Monotonically lowering $\bar{r}_t$ to promote peripheral-to-central ordering towards the start-point, this scheduling ensures a degradation level of $R'$ at $t=T_f$ assuming that the number of components removed per step is $n=\lfloor N_K(R'-1)/(R' T_f) \rfloor$. 

\textbf{Recovery operator.} Once random removal masks are determined, the relationship between $X_t$ and $X_0$ is given as: 
\begin{equation}
        \label{eq:removal_step}
        {X}_t =  \left( \prod_{\tau=1}^{t}{\mathbf{\Lambda}_{\tau} } \right) X_{0} = {\mathbf{\bar{\Lambda}}_{t}} X_{0}.
    \end{equation}
where ${\mathbf{\bar{\Lambda}}_{t}}$ is the cumulative frequency removal mask at $t$. The corresponding relationship in image domain is: 
\begin{align}
   & x_t  = \mathcal{F}^{-1} \big({\mathbf{\bar{\Lambda}}_{t}} \mathcal{F} (x_{0}) \big)= C_t x_{0},   \label{eq:gdp_exp1}
\end{align}
where $C_t$ is a corruption matrix in block Toeplitz form equivalent to the cumulative frequency removal at $t$ \cite{jain1989fundamentals}. We observe that Eq. \ref{eq:gdp_exp1} can be viewed as a special case of a generalized diffusion process with linear degradations \cite{daras2022soft}:
\begin{equation}
    \label{eq:gdp_exp2}
    x_t  = {\alpha_t} C_t x_{0} + {\sigma_t} z
\end{equation}
such that $\alpha_t=1$, $\sigma_t=0$. Based on this interpretation, FDB's training objective can be expressed via the following loss \cite{daras2022soft}:
\begin{eqnarray}
   L_{\mathrm{FDB}} = \mathbb{E}_{t, x_{(0,t)}} [\| C_t (G_{\theta}(x_t,t) - x_0) \|^2],
\end{eqnarray}
where $t\sim U(1,T_f)$. Since $\mathbf{\bar{\Lambda}}_{t}$ is diagonal with maximum eigenvalue 1, and $\mathcal{F}^{-1}$, $\mathcal{F}$ are orthonormal transformations:
\begin{equation}
   \| C_t \|=\|\mathcal{F}^{-1}{\mathbf{\bar{\Lambda}}_{t}} \mathcal{F}\|\leq \|\mathcal{F}^{-1}\| \|{\mathbf{\bar{\Lambda}}_{t}}\| \|\mathcal{F}\| = 1. 
\end{equation}
Further noting that $\| C_t (G_{\theta}(\cdot) - x_0) \|$$\,\leq\,$$\| C_t \| \|(G_{\theta}(\cdot) - x_0) \|$, FDB can be trained via a simplified upper-bound loss \cite{bansal2022cold}:
\begin{eqnarray}
   L_{\mathrm{FDB}-ub} = \mathbb{E}_{t, x_{(0,t)}} [\| (G_{\theta}(x_t,t) - x_0) \|^2].
   \label{eq:finalfdb}
\end{eqnarray}

\subsubsection{Sampling algorithm}
\label{sec:reverse}
Given $x_t$, the recovery operator in FDB estimates the completely dealiased image as $\Tilde{x}_0$$=$$G_{\theta}(x_t, t)$. The standard sampling algorithm for the generalized process in Eq. \ref{eq:gdp_exp2} can then be derived as \cite{song2022,daras2022soft}:
\begin{align}
        \label{eq:rev_fdb}
         {x}_{t-1} = x_t + (C_{t-1}-C_t)\Tilde{x}_0,
\end{align}
where $C_0=\mathrm{I}$, and the second term incrementally imputes frequency components between $t-1$ and $t$. To maintain stochasticity, here we employ independent random instances of $C_t$ during testing versus training. Note that the imputation term in Eq. \ref{eq:rev_fdb} performs hard dealiasing by only recovering $n$ frequencies selected in the set $S_t$, without altering the value of previously recovered frequencies in the aggregated set $S_{pre} = {\bigcup}_{\tau=T_{f}}^{t+1} S_{\tau}$. Assuming that the value of a frequency component $k$ is recovered at step $t_{k}$, it does not get updated during $t < t_{k}$ or $t>t_{k}$, so it is estimated only once. \textit{As such, the algorithm in Eq. \ref{eq:rev_fdb} would devoid FDB of the benefits of iterated Langevin sampling, compromising sample fidelity.}

\begin{algorithm}[t]
\small
    \caption{MRI reconstruction with FDB}
    \label{alg:sampling}
\KwIn{\\ $y$: MRI acquisition with acceleration rate $R$ \\ 
$A$: Imaging operator  \\
$G_{\theta}(x_t,t)$: Recovery operator \\ 
$T_f$: Number of diffusion steps \\
$R'$: Degradation level at start-point \\
$\bar{w}_t$: Correction weight at time step $t$ \\
} 
\KwOut{\\ $\hat{x}_0$: Reconstructed image  \\ {$\mbox{ }$}}
 \vspace{-0.15cm}
         $T_r = \lfloor T_f \frac{(R-1)R'}{(R'-1)R} \rfloor$ $\qquad$ $\triangleright$ number of reconstruction steps \\
         $\hat{x}_{T_r} = A^H y$ $\qquad \qquad$ $\triangleright$ zero-filled reconstruction \\
         $\bar{w}_{\{1,..,T_r\}} = \mathrm{resample}(w_{\{1,..,T_f\}},T_r/T_f)$ \\
         \For{$t = T_r, \hdots, 1$} {
             $\Tilde{x}_0 = G_{\theta}(\hat{x}_t, t)$ \qquad $\triangleright$ estimate `clean' image \\
             \eIf{$t \neq 1$}{
                 Sample $C_{t-1}$ by drawing $S_{t-1}$, given $C_t$ and $S_t$
                 }{
                 $C_{t-1}=\mathrm{I}$}
             $\dot{x}_{t-1} = \hat{x}_t + (C_{t-1}-C_t) \Tilde{x}_0 + \bar{w}_t  C_t (\Tilde{x}_0 - x_t)$\\
             $\hat{x}_{t-1} = \dot{x}_{t-1} + A^H (y - A \dot{x}_{t-1})$\\
         }
 \KwRet{$\hat{x}_0$}
\end{algorithm}

\begin{figure}[t]
        \centering
        \ifx\fastCompile\undefined
        \vspace{-0.2cm}
        \includegraphics[width=0.95\columnwidth]{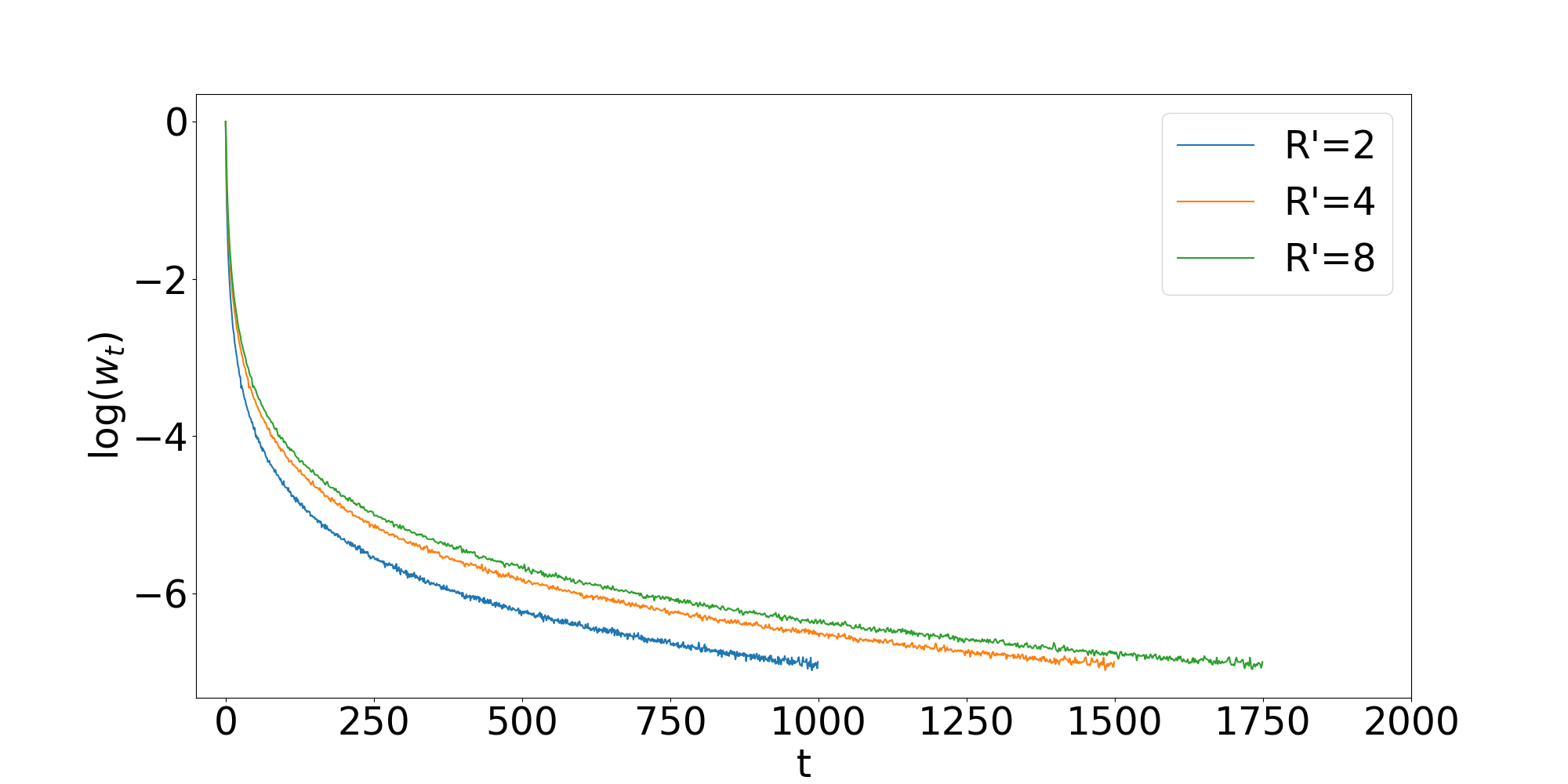}
        \captionsetup{justification=justified, singlelinecheck=false}
        \caption{The weighting parameter $w_t$ for the correction term in Eq. \ref{eq:rev_fdb_corr} was learned as described in Eq. \ref{eq:solutionw}. Results shown for $R'$=2 ($T_f$=1000), $R'$=4 ($T_f$=1500), and $R'$=8 ($T_f$=1750) on the IXI dataset.}
        \label{fig:w_t}
    \end{figure}

\textbf{Corrected sampling.} To mitigate this problem, we devise an enhanced sampling algorithm for FDB incorporating a correction term to achieve soft dealiasing during reverse steps: 
\begin{align}
        \label{eq:rev_fdb_corr}
         {x}_{t-1} = x_t + (C_{t-1}-C_t) \Tilde{x}_0 + w_t  C_t (\Tilde{x}_0 - x_t),
\end{align}
where $w_t \in [0 \mbox{ } 1]$ is a weighting parameter. The correction term examines the difference $(\Tilde{x}_0-x_t$) to estimate an update for frequency components recovered in previous steps $\{T_f,..,t\}$. Compared to earlier iterations (i.e., near $T_f$) where the set of recovered frequencies is compact, this correction becomes more critical in later iterations as the set grows in size. Thus, we adopted a learned schedule for $w_t$ to enhance recovery. Observing that $x_t=C_t x_t$, Eq. \ref{eq:rev_fdb_corr} can be rearranged:
\begin{align}
        \label{eq:rev_fdb_corr2}
         {x}_{t-1} = (C_{t-1}-C_t) \Tilde{x}_0 + C_t [w_t \Tilde{x}_0 + (1-w_t) x_t].
\end{align}
The second term in Eq. \ref{eq:rev_fdb_corr2} sets the values of previously recovered frequencies in $x_{t-1}$ to a weighted average of $\Tilde{x}_0$ and $x_t$. Assuming that $\Tilde{x}_0$ reasonably approximates $x_0$, $w_t$ can be learned via a regression problem in Fourier domain:
\begin{equation}
     \min_{w_t} \mbox{ } \mathbb{E}_{t,X_{(0,t-1,t)}} \underbrace{\big\| {X}_{t-1} - w_t {X}_0 - (1-w_t) X_t \big\|^2}_\text{$\Gamma$}.
     \label{eq:findingw}
\end{equation}
Since $X_t^T X_0=(\mathbf{\bar{\Lambda}}_{t} X_0)^T (\mathbf{\bar{\Lambda}}_{t} X_0)=\Vert X_t \Vert^2$, $\Gamma$ is given as:
\begin{align}
     \Gamma = & \| {X}_{t-1} \|^2 + w_t^2 \|{X}_0\|^2 + (1-w_t)^2 \|X_t\|^2 - 2w_t \|X_{t-1}\|^2 \nonumber \\
     & - 2(1-w_t)\|X_{t}\|^2 + 2w_t(1-w_t)\|X_{t}\|^2, \\
     = & w_t^2 \left( \|{X}_0\|^2 - \|X_t\|^2 \right) - 2 w_t \left( \| {X}_{t-1} \|^2 - \|X_t\|^2 \right) \nonumber \\
    & + \left( \| {X}_{t-1} \| - \|X_t\|^2 \right).
\end{align}
Note that the partial derivative of $\Gamma$ with respect to $w_t$ is:
\begin{align}
  {\partial \Gamma}/{\partial w_t} =   2 w_t \left( \|{X}_0\|^2 - \|X_t\|^2 \right) - 2 \left( \| {X}_{t-1} \|^2 - \|X_t\|^2 \right). 
\end{align}
Thus, the analytical solution to Eq. \ref{eq:findingw} can be expressed as:
\begin{align}
    w_t = \frac{ \mathbb{E}_{t,X_{(t-1,t)}} \left( \|{X}_{t-1}\|^2 - \|X_t\|^2 \right)}{\mathbb{E}_{t,X_{(0,t)}} \left( \| {X}_{0} \|^2 - \|X_t\|^2 \right)}, \label{eq:solutionw}
\end{align}
where the numerator and denominator reflect the expected energy differences between $X_{t-1}$, $X_t$ and $X_0$, $X_t$, respectively, which can be estimated via a simple Monte-Carlo simulation over the training set. Note that $w_t$ takes a maximum value of 1 attained at $t=1$. Assuming that k-space energy distribution of MR images abide by an approximate inverse-power law (i.e., $\propto 1/r^{\nu}$), $w_t$ should show exponential behavior \cite{bialek1994}. Indeed, we find that $w_t$ follows an exponential trend rising from $t=T_f$ to $t=1$ (Fig. \ref{fig:w_t}).  

\textbf{MRI reconstruction.} Reconstruction is achieved by interleaving reverse diffusion steps for FDB with data-consistency projections (Alg. \ref{alg:sampling}). The trained FDB prior maps $R'$-fold undersampled data onto fully-sampled data in $T_f$ steps, by recovering $n \approx N_K(R'-1)/(R' T_f)$ frequency components in each step. Thus, to reconstruct an $R$-fold undersampled acquisition with $N_K(R-1)/R$ missing components, diffusion sampling with the FDB prior must last for $T_r = \lfloor T_f \frac{(R-1)R'}{(R'-1)R} \rfloor$ steps. As sinusoidal time encoding is used in FDB, sampling can be extrapolated to a broader time range (i.e., $T_r > T_f$) to account for cases where $R>R'$ \cite{nichol2021improved}. Yet, $w_t$ is resampled across time to $\bar{w}_t$ to cover the designated range $[w_{1} \mbox{ } w_{T_f}]$ in $T_r$ steps. Sampling is initiated at $T_r$ with the least-squares reconstruction of undersampled data, $\hat{x}_{T_r}=A^{H}y$ \cite{chung2022cvpr}. The mappings at time step $t$ can then be expressed as: 
\begin{align}
        \label{eq:recon_fdb}
         & {\dot{x}}_{t-1}\, = \,\, \hat{x}_t + (C_{t-1}-C_t) \Tilde{x}_0 + \bar{w}_t  C_t (\Tilde{x}_0 - x_t), \\
        & \hat{x}_{t-1} = \dot{x}_{t-1} + A^H (y - A \dot{x}_{t-1}).
\end{align}
Finally, $\hat{x}_0$ is taken as the reconstructed image.

\section{Methods}

\subsection{Datasets}

    Experiments were performed on brain MRI data from IXI\footnote{https://brain-development.org/ixi-dataset/} and fastMRI \cite{fastmri}. Subjects were split into non-overlapping training, validation, test sets. IXI contained coil-combined magnitude images for T\textsubscript{1}, T\textsubscript{2} and PD contrasts, which were analyzed as single-coil acquisitions. A subject-level split of (21,15,30) was used. fastMRI contained multi-coil complex k-space data for T\textsubscript{1}, T\textsubscript{2} and FLAIR contrasts. Geometric coil compression was used to derive 5 virtual coils that preserved over 90\% of the variance in the original data \cite{zhang2013coil}. A subject-level split of (240,60,120) was used. Retrospective undersampling was used based on a normal density across the two-dimensional transverse plane at acceleration rates $R= 4-8$ \cite{lustig2007sparse}. To construct the imaging operators for reconstruction, coil sensitivities were estimated using ESPIRiT on a central calibration region \cite{uecker2014espirit}. Volumetric k-space data were inverse Fourier transformed and split across the readout dimension, and each cross-section was reconstructed individually.

\subsection{Competing Methods}
    FDB was demonstrated against 7 state-of-the-art methods from the literature. Network models used two separate input and output channels to represent real and imaginary components. Diffusion models based on noise addition used exponential scheduling with $(\beta_{min}, \beta_{max}) = (0.1, 20)$ \cite{song2022}. For each method, key hyperparameters were selected to maximize performance on the validation set. Modeling was performed via PyTorch on an Nvidia RTX 3090. Models were trained using the Adam optimizer with $(\beta_1, \beta_2) = (0.5, 0.9)$.

    \textbf{FDB}: FDB used the architecture in \cite{ho2020denoising}. Hyperparameters were set as $\eta=10^{-4}$ learning rate, $T_f$=1000, $E$=40 epochs for IXI, $E$=10 for fastMRI, and $R'$=2. 
    
    \textbf{LORAKS}: A low-rank reconstruction was considered \cite{haldar2016p}. The k-space neighborhood radius and the rank of the system matrix were set as (2,6) for IXI, and (2,30) for fastMRI.

    \textbf{D5C5}: A task-specific unrolled model was considered \cite{Schlemper2017}. Hyperparameters were $\eta=2 \times 10^{-4}$, $E$=100.

    \textbf{rGAN}: A task-specific GAN model was considered \cite{dar2020prior}. Hyperparameters were set as $\eta=2 \times 10^{-4}$, $E$=100, adversarial and pixel-wise loss weights of (1,100).

    \textbf{DDPM}: A task-agnostic diffusion model was implemented \cite{ho2020denoising}. Hyperparameters were set as $\eta=10^{-4}$, $T$=1000, $E$=40.

    \textbf{CDiffMR}: A cold diffusion model was considered based on an asymptotic start-point derived via k-space undersampling \cite{huang2023cdiffmr}. Reconstructions were initiated via the zero-filled (ZF) reconstruction at the time step corresponding to $R$ of the test acquisition \cite{huang2023cdiffmr}. Hyperparameters were set as $\eta=10^{-4}$, $T$=1000, $E$=40.  

    \textbf{I\SP{2}SB}: An unconstrained diffusion bridge with start- and end-point identical to FDB was considered \cite{liu2023i2sb}. Reconstructions were initiated via the ZF reconstruction of the test acquisition. Hyperparameters were $\eta=10^{-4}$, $T$=200, $E$=40. 
        
    \textbf{DB\SB{blur}}: A constrained diffusion bridge based on a blur operator was considered \cite{fabian2023diracdiffusion}. The blur operator used a two-dimensional Gaussian kernel of size 15 and standard deviation 0.01 \cite{daras2022soft}. Reconstructions were initiated via the ZF reconstruction of the test acquisition. Hyperparameters were $\eta=10^{-4}$, $T$=200, $E$=40.

 \begin{figure*}[t]
        \centering
        \ifx\fastCompile\undefined
        \includegraphics[width=\textwidth]{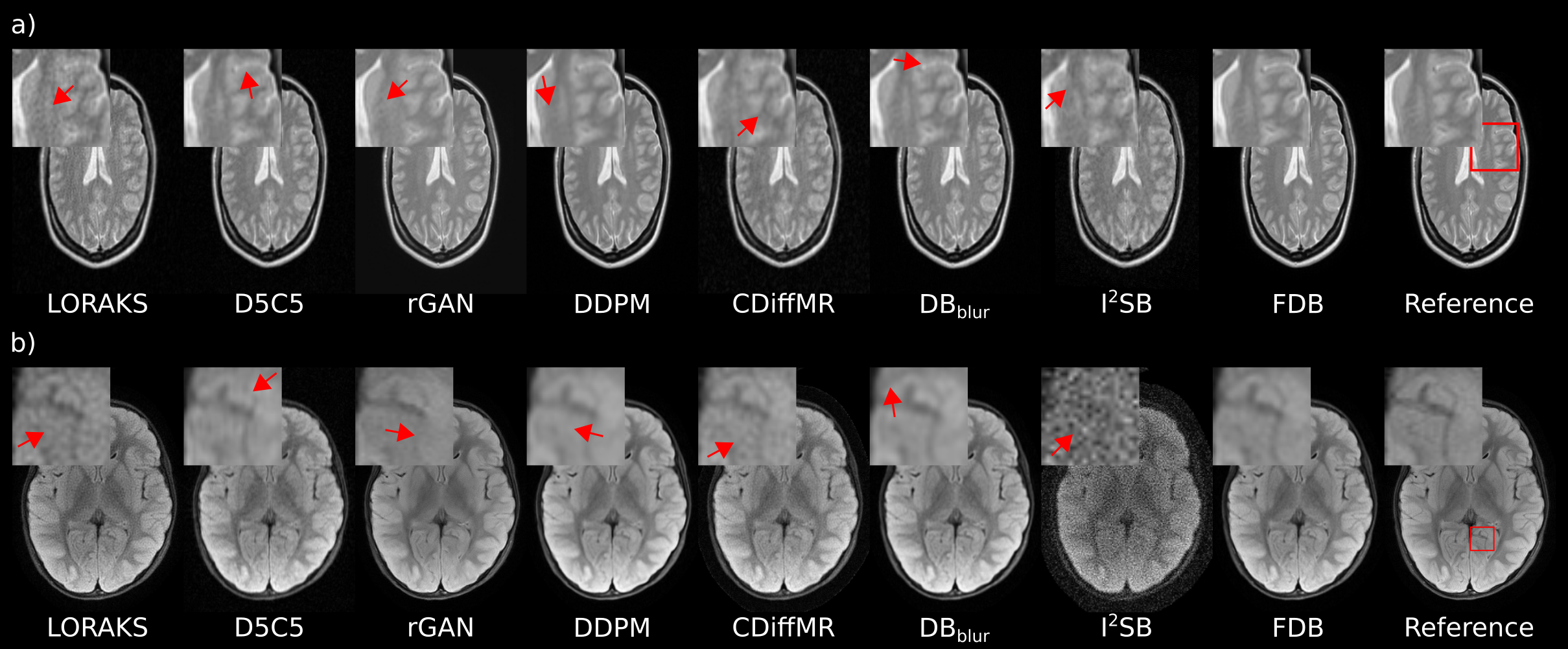}
        \captionsetup{justification=justified, singlelinecheck=false}
        \caption{Within-domain reconstructions of representative cross-sections from (a) a PD acquisition in IXI at $R=4$ and (b) a FLAIR acquisition in fastMRI at $R=8$. Images from competing methods are displayed along with the reference image computed via Fourier reconstruction of fully-sampled data. Zoom-in windows are included to highlight differences among methods, and the red box on the reference image marks the zoomed region.}
        \label{fig:within-domain}
   \end{figure*}
    
\subsection{Analyses}
For each dataset, model training was performed on data aggregated across multiple contrasts (T\SB{1}, T\SB{2}, PD in IXI and T\SB{1}, T\SB{2}, FLAIR in fastMRI). No explicit contrast information was provided during training, and data samples were randomly drawn for the aggregated training set. Task-specific models were trained for each $R$ separately to maintain high performance. Meanwhile, task-agnostic models were trained without knowledge of $R$. Models were trained and tested on two-dimensional cross sections. Single-coil reconstructions were conducted on IXI, and multi-coil reconstructions were conducted on fastMRI. Reconstruction performance was assessed by quantifying peak signal-to-noise ratio (PSNR) and structural similarity index (SSIM) between recovered and reference images. Reference images were derived via Fourier reconstruction of fully-sampled acquisitions. Significance of performance differences among competing methods were assessed via non-parametric Wilcoxon signed-rank tests.

\section{Results}

\subsection{Within-Domain Reconstruction}

    FDB was demonstrated against LORAKS (traditional method), D5C5 (task-specific unrolled model), rGAN (task-specific adversarial model), DDPM (task-agnostic diffusion prior), CDiffMR (cold diffusion prior based on undersampling), I\SP{2}SB (unconstrained diffusion bridge), and DB\textsubscript{blur} (blur-constrained diffusion bridge). First, within-domain reconstruction tasks were examined where models were trained and tested on the same dataset and sampling density. Note that all competing methods except LORAKS were trained, so their data-driven priors naturally depend on the dataset used. Meanwhile, DDPM and DB\SB{blur} do not use information regarding the imaging operator and thereby the sampling density during training. Yet, network inputs to D5C5 and rGAN depend on the undersampling patterns used in the training set, and start-points of the diffusion process in CDiffMR, I\SP{2}SB and FDB are obtained by assuming a certain sampling density. 
    
    Performance metrics are listed in Table \ref{tab:singlecoil} for IXI, and in Table \ref{tab:multicoil} for fastMRI. Overall, FDB is the top performer in all examined tasks and datasets (p$<$0.05). On average across tasks, FDB outperforms LORAKS by 7.5dB PSNR, 24.4\% SSIM, task-specific priors by 4.3dB PSNR and 12.0\% SSIM, diffusion priors by 4.8dB and 7.6\% SSIM, and diffusion bridges by 6.5dB PSNR and 9.8\% SSIM. Compared against the second-best method in each task, FDB achieves an average improvement of 2.8dB PSNR and 1.1\% SSIM.     
    Representative reconstructions are depicted in Fig. \ref{fig:within-domain}a for IXI, and Fig. \ref{fig:within-domain}b for fastMRI. In general, LORAKS, CDiffMR, and I\SP{2}SB show high noise amplification and artifacts; rGAN shows occasional residual artifacts and moderate losses in structural details; D5C5, DDPM, and DB\textsubscript{blur} show a visible degree of blurring due to oversmoothing. In contrast, FDB yields images with more accurate depiction of tissue structure along with lower artifacts and noise than competing methods. Taken together, these results suggest that the increased alignment between the transformation captured by FDB and the target transformation for MRI reconstruction enables FDB to achieve superior performance against common diffusion priors and diffusion bridges based on task-irrelevant degradation operators.

   \begin{table}[t]
        \centering
        \captionsetup{justification=justified,singlelinecheck=false}
        \caption{Within-domain performance on IXI at $R= 4,8$. Results averaged across T\SB{1}, T\SB{2}, PD contrasts. PSNR (dB) and SSIM (\%) listed as mean$\pm$std across the test set. Boldface marks the top performing method.}
        \resizebox{0.75\columnwidth}{!}{%
            \begin{tabular}{|l|c|c|c|c|}
                \cline{2-5}
                \multicolumn{1}{c|}{} & \multicolumn{2}{|c|}{$R = 4$} & \multicolumn{2}{|c|}{$R = 8$} \\
                \cline{2-5}
                \multicolumn{1}{c|}{} & PSNR & SSIM & PSNR & SSIM \\
                \hline
                LORAKS & 31.3$\pm$1.7 & 63.2$\pm$4.4 & 27.9$\pm$1.3 & 55.7$\pm$4.1 \\
                \hline
                D5C5 & 33.8$\pm$0.6 & 78.3$\pm$1.6 & 30.0$\pm$0.6 & 66.7$\pm$1.9 \\
                \hline
                rGAN & 36.6$\pm$3.0 & 85.1$\pm$7.1 & 32.6$\pm$2.9 & 79.5$\pm$7.3 \\
                \hline
                DDPM & 38.0$\pm$2.5 & 97.6$\pm$0.7 & 32.9$\pm$2.6 & 95.4$\pm$1.4 \\
                \hline
                CDiffMR & 32.4$\pm$2.4 & 86.0$\pm$2.2 & 29.3$\pm$2.4 & 80.7$\pm$2.6 \\
                \hline
                I\textsuperscript{2}SB & 30.3$\pm$2.6 & 83.0$\pm$2.8 & 28.2$\pm$2.6 & 79.9$\pm$3.1 \\
                \hline
                DB\textsubscript{blur} & 36.7$\pm$2.6 & 96.7$\pm$0.9 & 33.4$\pm$2.7 & 95.3$\pm$1.2 \\
                \hline
                 FDB & \textbf{43.9$\pm$3.0} & \textbf{99.3$\pm$0.9} & \textbf{35.7$\pm$2.7} & \textbf{97.1$\pm$1.1} \\
                \hline
            \end{tabular}
        }
        \label{tab:singlecoil}
    \end{table}
  \begin{table}[t]
        \centering
        \captionsetup{justification=justified,singlelinecheck=false}
        \caption{Within-domain performance on fastMRI at $R= 4, 8$. Results averaged across T\SB{1}, T\SB{2}, FLAIR contrasts.}
        \resizebox{0.75\columnwidth}{!}{%
            \begin{tabular}{|l|c|c|c|c|}
                \cline{2-5}
                \multicolumn{1}{c|}{} & \multicolumn{2}{|c|}{$R = 4$} & \multicolumn{2}{|c|}{$R = 8$} \\
                \cline{2-5}
                \multicolumn{1}{c|}{} & PSNR & SSIM & PSNR & SSIM \\
                \hline
                LORAKS & 31.7$\pm$2.6 & 82.6$\pm$7.7 & 31.0$\pm$2.4 & 83.2$\pm$8.5 \\
                \hline
                D5C5 & 35.3$\pm$1.5 & 90.5$\pm$2.3 & 32.2$\pm$1.4 & 85.0$\pm$3.1 \\
                \hline
                rGAN & 35.9$\pm$2.8 & 93.1$\pm$4.7 & 32.9$\pm$2.8 & 90.1$\pm$5.9 \\
                \hline
                DDPM & 36.1$\pm$2.3 & 93.7$\pm$4.6 & 34.0$\pm$2.3 & 91.1$\pm$5.6 \\
                \hline
                CDiffMR & 31.4$\pm$2.7 & 78.5$\pm$8.1 & 31.2$\pm$2.3 & 80.5$\pm$7.4 \\
                \hline
                I\textsuperscript{2}SB & 28.4$\pm$2.1 & 75.7$\pm$4.5 & 26.2$\pm$2.1 & 70.9$\pm$5.0 \\
                \hline
                DB\textsubscript{blur} & 35.6$\pm$2.3 & 93.6$\pm$4.5 & 33.5$\pm$2.3 & 90.9$\pm$5.6 \\
                \hline
                FDB & \textbf{37.4$\pm$2.5} & \textbf{94.3$\pm$4.5} & \textbf{35.0$\pm$2.4} & \textbf{91.5$\pm$5.6} \\
                \hline
            \end{tabular}
        }
        \label{tab:multicoil}
    \end{table}

    \begin{figure*}[t]
        \centering
        \ifx\fastCompile\undefined
        \includegraphics[width=\textwidth]{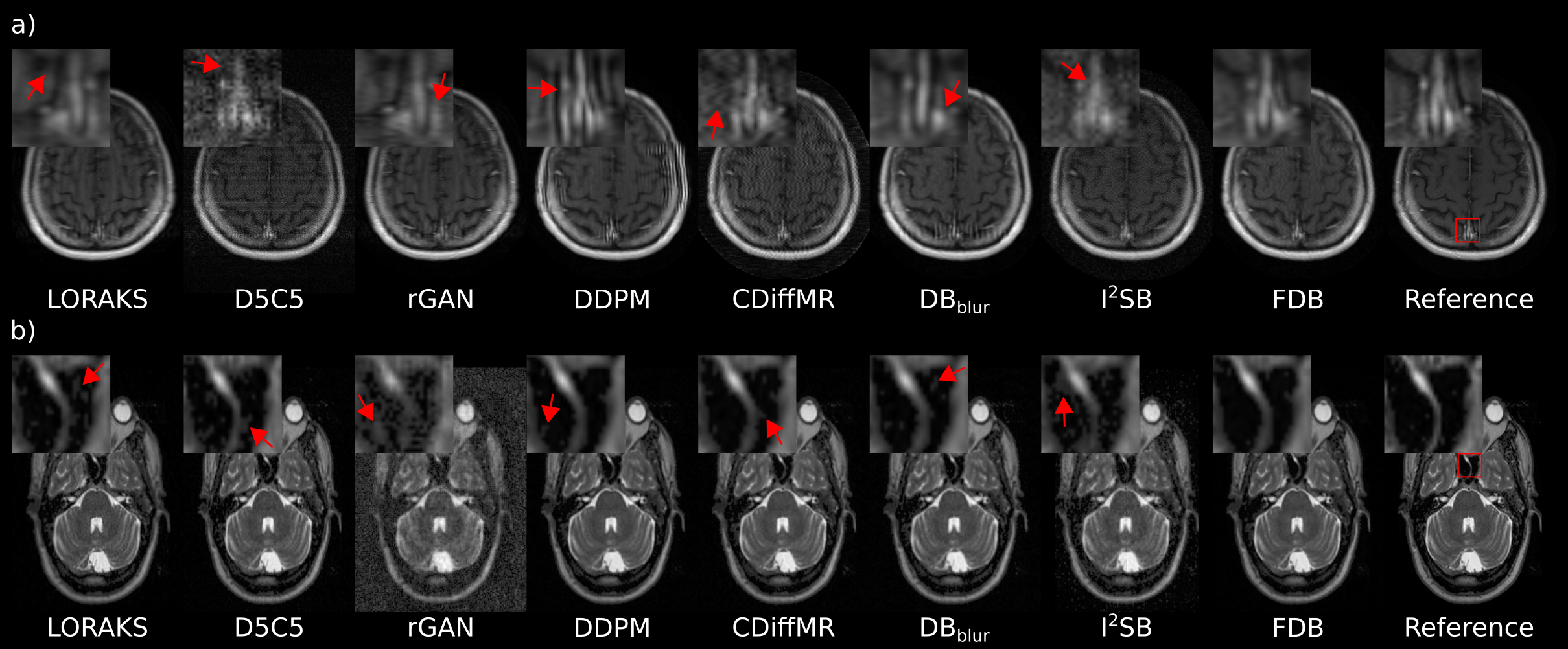}
        \captionsetup{justification=justified, singlelinecheck=false}
        \caption{Cross-domain reconstructions of representative cross-sections. \textbf{(a)} A T\SB{1} acquisition in fastMRI at $R=8$ is shown. Models trained on 2D k-space sampling patterns were tested on 1D patterns under matched $R$. \textbf{(b)} A T\SB{2} acquisition in IXI at $R=8$ is shown. Models trained on multi-coil fastMRI data were tested on single-coil IXI data. Images from competing methods are displayed along with the reference image computed via Fourier reconstruction of fully-sampled data. Zoom-in windows are included to highlight differences among methods, and the red box on the reference image marks the zoomed region.}
        \label{fig:cross-domain}
    \end{figure*}

    \begin{table}[t]
        \centering
        \captionsetup{justification=justified,singlelinecheck=false}
        \caption{Cross-domain performance for models trained on 2D k-space sampling patterns, tested on 1D patterns for IXI at $R= 4, 8$.}
        \resizebox{0.75\columnwidth}{!}{%
            \begin{tabular}{|l|c|c|c|c|}
                \cline{2-5}
                \multicolumn{1}{c|}{} & \multicolumn{2}{|c|}{$R = 4$} & \multicolumn{2}{|c|}{$R = 8$} \\
                \cline{2-5}
                \multicolumn{1}{c|}{} & PSNR & SSIM & PSNR & SSIM \\
                \hline
                LORAKS & 24.3$\pm$0.9 & 42.0$\pm$2.6 & 21.3$\pm$0.7 & 34.2$\pm$2.7 \\
                \hline
                D5C5 & 24.4$\pm$0.7 & 55.0$\pm$3.7 & 22.7$\pm$0.6 & 49.8$\pm$4.0 \\
                \hline
                rGAN & 29.3$\pm$2.4 & 49.1$\pm$6.1 & 25.8$\pm$2.3 & 39.2$\pm$5.7 \\
                \hline
                DDPM & 30.1$\pm$2.8 & 94.3$\pm$1.9 & 25.1$\pm$2.5 & 85.5$\pm$3.6 \\
                \hline
                CDiffMR & 28.6$\pm$2.6 & 83.7$\pm$2.5 & 24.8$\pm$2.5 & 74.2$\pm$3.6 \\
                \hline
                I\textsuperscript{2}SB & 27.8$\pm$2.8 & 85.0$\pm$3.2 & 25.5$\pm$2.7 & 81.0$\pm$4.2 \\
                \hline
                DB\textsubscript{blur} & 31.8$\pm$2.9 & 94.8$\pm$1.6 & 27.0$\pm$2.9 & 87.5$\pm$3.2 \\
                \hline
                FDB & \textbf{34.0$\pm$2.9} & \textbf{96.7$\pm$1.2} & \textbf{28.1$\pm$2.7} & \textbf{89.8$\pm$2.7} \\
                \hline
            \end{tabular}
        }
        \label{tab:2dvd_to_1dvd_ixi}
    \end{table}

    \begin{table}[t]
        \centering
        \captionsetup{justification=justified,singlelinecheck=false}
        \caption{Cross-domain performance for models trained on 2D k-space sampling patterns, tested on 1D patterns for fastMRI at $R= 4, 8$.}
        \resizebox{0.75\columnwidth}{!}{%
            \begin{tabular}{|l|c|c|c|c|}
                \cline{2-5}
                \multicolumn{1}{c|}{} & \multicolumn{2}{|c|}{$R = 4$} & \multicolumn{2}{|c|}{$R = 8$} \\
                \cline{2-5}
                \multicolumn{1}{c|}{} & PSNR & SSIM & PSNR & SSIM \\
                \hline
                LORAKS & 28.1$\pm$2.3 & 85.3$\pm$5.6 & 24.5$\pm$2.2 & 76.9$\pm$6.5 \\
                \hline
                D5C5 & 26.4$\pm$1.6 & 64.0$\pm$8.0 & 24.7$\pm$9.6 & 53.3$\pm$4.0 \\
                \hline
                rGAN & 29.1$\pm$2.9 & 86.1$\pm$5.3 & 25.4$\pm$2.6 & 77.3$\pm$6.8 \\
                \hline
                DDPM & 29.4$\pm$2.0 & 88.8$\pm$5.5 & 25.7$\pm$1.5 & 80.6$\pm$6.5 \\
                \hline
                CDiffMR & 27.1$\pm$2.2 & 68.7$\pm$9.2 & 25.1$\pm$2.1 & 63.7$\pm$9.7 \\
                \hline
                I\textsuperscript{2}SB & 24.3$\pm$4.5 & 68.6$\pm$8.8 & 21.3$\pm$5.2 & 63.3$\pm$9.5 \\
                \hline
                DB\textsubscript{blur} & 30.2$\pm$2.1 & 89.2$\pm$5.8 & 27.7$\pm$2.0 & 82.0$\pm$7.5 \\
                \hline
                FDB & \textbf{30.6$\pm$2.2} & \textbf{89.5$\pm$6.0} & \textbf{27.8$\pm$2.0} & \textbf{82.7$\pm$7.3} \\
                \hline
            \end{tabular}
        }
        \label{tab:2dvd_to_1dvd_fastmri}
    \end{table}
 
\subsection{Cross-Domain Reconstruction}
    Next, cross-domain reconstruction tasks involving shifts in sampling density or dataset between training and test sets were examined. For shifts in sampling density, generalization from from 2D to 1D variable-density sampling was considered under matching acceleration rate $R$. Cross-domain performance metrics listed in Tables \ref{tab:2dvd_to_1dvd_ixi} and \ref{tab:2dvd_to_1dvd_fastmri} indicate that FDB is the top performer in all tasks (p$<$0.05), except for DB\SB{blur} that yields similar PSNR on fastMRI at $R=8$. On average across tasks, FDB outperforms LORAKS by 5.6dB PSNR, 30.1\% SSIM, task-specific priors by 4.2dB PSNR and 30.5\% SSIM, diffusion priors by 3.1dB and 9.7\% SSIM, and diffusion bridges by 3.2dB PSNR and 8.3\% SSIM. Over the second-best method in each task, FDB achieves an average improvement of 1.0dB PSNR and 1.3\% SSIM. For shifts in dataset, generalization from the multi-coil fastMRI dataset to single-coil IXI dataset was examined as listed in Table \ref{tab:multicoil_to_singlecoil}. Again, FDB achieves the highest performance among competing methods (p$<$0.05). On average across tasks, FDB outperforms LORAKS by 4.8dB PSNR, 34.8\% SSIM, task-specific priors by 8.4dB PSNR and 45.0\% SSIM, diffusion priors by 2.9dB and 7.6\% SSIM, and diffusion bridges by 3.0dB PSNR and 7.0\% SSIM. Over the second-best method in each task, FDB achieves an average improvement of 0.9dB PSNR and 1.4\% SSIM. Taken together, these results suggest that FDB captures a more reliable prior against domain shifts in sampling density and data distribution than competing methods, including diffusion priors.
    
    Representative images from competing methods are depicted in Fig. \ref{fig:cross-domain}a for domain shifts in sampling density, and Fig. \ref{fig:cross-domain}b for domain shifts in dataset. In general, LORAKS shows a degree of residual artifacts; CDiffMR and I\SP{2}SB show high noise amplification and artifacts; DDPM and DB\textsubscript{blur} show occasional ringing artifacts and a degree of blur. As expected, task-specific priors D5C5 and rGAN yield notable residual artifacts and noise amplification in cross-domain settings. Meanwhile, FDB produces more accurate depiction of tissue structure, and relatively low artifacts and noise than baselines.

\subsection{Ablation Studies}
    Ablation studies were conducted to examine the influence of major design parameters in FDB to reconstruction performance. FDB captures a delimited transformation between an end-point of fully-sampled data and a start-point of moderately undersampled data, rendering the degradation level $R'$ at the start-point critical. Performance for FDB variants trained using $R'$=2-16 are listed in Table \ref{tab:kspace-ablation-singlecoil}. In general, employing lower $R'$ values yields consistently higher performance across $R$ prescribed for test acquisitions. This finding suggests that it is not ideal to strictly match $R'$ and $R$, and thereby helps explain the superior performance of FDB against diffusion priors with asymptotic start-points (e.g., DDPM and CDiffMR).

    \begin{table}[t]
        \centering
        \captionsetup{justification=justified,singlelinecheck=false}
        \caption{Cross-domain performance for models trained on multi-coil fastMRI data, and tested on single-coil IXI data at $R= 4, 8$.}
        \resizebox{0.75\columnwidth}{!}{%
            \begin{tabular}{|l|c|c|c|c|}
                \cline{2-5}
                \multicolumn{1}{c|}{} & \multicolumn{2}{|c|}{$R = 4$} & \multicolumn{2}{|c|}{$R = 8$} \\
                \cline{2-5}
                \multicolumn{1}{c|}{} & PSNR & SSIM & PSNR & SSIM \\
                \hline
                LORAKS & 31.3$\pm$1.7 & 63.2$\pm$4.4 & 27.9$\pm$1.3 & 55.7$\pm$4.1 \\
                \hline
                D5C5 & 34.4$\pm$0.7 & 80.3$\pm$1.6 & 30.3$\pm$0.6 & 68.1$\pm$2.0 \\
                \hline
                rGAN & 19.9$\pm$3.5 & 25.1$\pm$7.7 & 19.2$\pm$3.1 & 23.5$\pm$6.7 \\
                \hline
                DDPM & 34.8$\pm$2.7 & 93.6$\pm$1.4 & 32.2$\pm$2.6 & 92.0$\pm$1.8 \\
                \hline
                CDiffMR & 30.6$\pm$2.7 & 82.7$\pm$3.0 & 28.2$\pm$2.6 & 78.2$\pm$3.3 \\
                \hline
                I\textsuperscript{2}SB & 31.2$\pm$2.5 & 84.6$\pm$2.5 & 28.9$\pm$2.6 & 81.2$\pm$2.8 \\
                \hline
                DB\textsubscript{blur} & 33.8$\pm$2.8 & 92.0$\pm$1.8 & 31.5$\pm$2.8 & 91.1$\pm$2.0 \\
                \hline
                FDB & \textbf{36.0$\pm$2.7} & \textbf{95.4$\pm$1.0} & \textbf{32.7$\pm$2.7} & \textbf{93.0$\pm$1.3} \\
                \hline
            \end{tabular}
        }
        \label{tab:multicoil_to_singlecoil}
    \end{table}

    We also examined the influence of main parameters related to the diffusion process and sampling algorithm. FDB's diffusion process uses a k-space radius threshold to promote peripheral-to-central ordering and variable-density frequency removal, removes a fixed number of (i.e., n) frequency components at each time step, and progressively degrades fully-sampled data via frequency removal to obtain intermediate samples $X_t$. To assess the importance of these parameters, a `$C_t$ - uniform'  variant was formed that ablated the radius threshold to follow uniform-density frequency removal without ordering, a `n - log schedule' variant was formed where the number of frequency components selected for removal was varied on a log schedule across time, and a `$X_t$ - averaging constraint' variant was formed where mean of intermediate samples were expressed as a weighted linear average of start- and end-points based on temporal distance \cite{delbracio2023inversion}.  
    
    Meanwhile, FDB's sampling algorithm prescribes independent instances of degradation operators $C_t$ between training and test sets, includes a correction term for soft dealiasing, uses a learned schedule for the correction-term weight $w_t$. To assess the importance of these parameters, `$C_t$ - fixed' variant was formed that used shared instances of $C_t$ between training-test sets, a `w/o correction' variant was formed by ablating the correction term, and a `$w_t$ - linear schedule' variant was formed by varying $w_t$ linearly across time. Table \ref{tab:inference-ablation} lists performance metrics for all FDB variants. The proposed FDB parameters yield generally superior performance against variants, corroborating the importance of these design choices.

\section{Discussion}
The proposed FDB prior was demonstrated against prominent deep-learning baselines for MRI reconstruction including conditional models that capture task-specific priors and unconditional models that capture task-agnostic priors. For both within- and cross-domain settings, we find that FDB offers enhanced image quality against baselines for the examined acceleration rates, tissue contrasts, and datasets. The relatively higher benefits of FDB over task-specific priors in cross-domain reconstruction corroborate recent findings that generative modeling approaches can help improve generalization in MRI reconstruction \cite{jalaln2021nips,chung2022cvpr}. Yet, the elevated performance of FDB over common diffusion priors and diffusion bridges based on task-irrelevant degradations suggest that utilization of task-relevant degradations that respect the physics of accelerated MRI serves a critical role in obtaining high-fidelity image priors during generative modeling. 

Conventional inference in diffusion models employs Langevin sampling across thousands of steps to generate images, resulting in characteristically slow inference. Unlike diffusion priors with asymptotic normality assumptions that initiate sampling with a random noise image, FDB initiates sampling with the least-squares reconstruction of undersampled acquisitions, recently shown to improve efficiency \cite{chung2022cvpr}. Yet, FDB has longer inference times compared to other generative models such as adversarial priors with one-step image sampling. Several acceleration strategies can improve efficiency in FDB. A training-phase approach is to integrate adversarial learning to enable reverse diffusion over large step sizes \cite{gungor2022adaptive}. Alternatively, inference-phase approaches can be adopted to perform interleaved sampling followed by a refinement procedure \cite{peng2022towards}, or via distillation of the trained priors \cite{bedel2023dreamr}. Future studies are warranted to assess the influence of these acceleration approaches on FDB's performance.

    \begin{table}[t]
        \centering
        \captionsetup{justification=justified,singlelinecheck=false}
        \caption{Performance of FDB priors on IXI. $R' = 2-16$ is the acceleration rate equivalent to the degradation at the start-point of the diffusion process during training, and $R= 4, 8$ is the acceleration rate during testing.}
        \resizebox{0.72\columnwidth}{!}{%
            \begin{tabular}{|c|c|c|c|c|}
                \cline{2-5}
                \multicolumn{1}{c|}{} & \multicolumn{2}{|c|}{$R = 4$} & \multicolumn{2}{|c|}{$R = 8$} \\
                \cline{2-5}
                \multicolumn{1}{c|}{} & PSNR & SSIM & PSNR & SSIM \\
                \hline
                $R' = 2$ & \textbf{43.7$\pm$2.9} & \textbf{99.3$\pm$0.9} & \textbf{36.9$\pm$2.6} & \textbf{97.8$\pm$1.0} \\
                \hline
                $R' = 4$ & 42.5$\pm$2.9 & 99.1$\pm$0.7 & 35.0$\pm$2.7 & 96.7$\pm$1.0 \\
                \hline
                $R' = 8$ & 42.6$\pm$2.8 & 99.2$\pm$0.5 & 34.6$\pm$2.5 & 96.5$\pm$1.0 \\
                \hline
                 $R' = 16$ & 42.4$\pm$3.1 & 99.1$\pm$0.6 & 34.6$\pm$2.6 & 96.6$\pm$1.0 \\
                \hline
            \end{tabular}
        }
        \label{tab:kspace-ablation-singlecoil}
    \end{table}

    \begin{table}[t]
        \centering
        \captionsetup{justification=justified,singlelinecheck=false}
        \caption{Performance of FDB variants on IXI at $R=4,8$. Several variants concerned ablations on the diffusion process;`$C_t$ - uniform': radius threshold ablated to follow a uniform density for frequency removal, `n - log schedule': number of removed frequency components varied on a log schedule, `$X_t$ - averaging constraint': mean for samples at intermediate steps expressed as a linear average of undersampled and fully-sampled data. Several variants concerned ablations on the sampling algorithm: `$C_t$ - fixed': shared instances of degradation operators used between training-test sets, `w/o correction': correction term ablated, `$w_t$ - linear schedule': the weight for the correction term was varied on a linear schedule, }
        \resizebox{0.99\columnwidth}{!}{%
            \begin{tabular}{|c|c|c|c|c|c|}
                \cline{3-6}
                \multicolumn{2}{c|}{} & \multicolumn{2}{|c|}{$R = 4$} & \multicolumn{2}{|c|}{$R = 8$} \\
                \cline{3-6}
                \multicolumn{2}{c|}{} & PSNR & SSIM & PSNR & SSIM \\
                \hline
                & FDB & \textbf{43.7$\pm$2.9} & \textbf{99.3$\pm$0.9} & \textbf{36.9$\pm$2.6} & 97.8$\pm$1.0 \\
                \hline
                \multirow{3}{*}{\rotatebox[origin=c]{90}{\textbf{Process}}} & $C_t$ - uniform & 43.0$\pm$2.7 & \textbf{99.3$\pm$1.0} & 36.8$\pm$2.5 & \textbf{98.1$\pm$1.0} \\
                \cline{2-6}
                & n - log schedule & 40.7$\pm$3.0 & 98.7$\pm$2.9 & 35.5$\pm$2.8 & 97.3$\pm$0.9 \\
                \cline{2-6}
                & $X_t$ - averaging constraint & 17.0$\pm$2.8 & 50.9$\pm$5.2 & 17.1$\pm$2.6 & 51.4$\pm$4.9 \\
                \hline
                \multirow{3}{*}{\rotatebox[origin=c]{90}{\textbf{Sampling}}} & $C_t$ - fixed & 43.4$\pm$2.9 & \textbf{99.3$\pm$0.8} & 36.2$\pm$2.5 & 97.4$\pm$1.1 \\
                \cline{2-6}
                & w/o correction & 39.0$\pm$2.6 & 96.5$\pm$0.8 & 34.2$\pm$2.6 & 94.0$\pm$1.3 \\
                \cline{2-6}
                & $w_t$ - linear schedule & 24.4$\pm$2.4 & 85.1$\pm$3.1 & 20.8$\pm$2.4 & 80.2$\pm$2.7 \\
                \hline
            \end{tabular}
        }
        \label{tab:inference-ablation}
    \end{table}

We find that FDB shows better generalization than baselines under a fair level of domain shifts in k-space sampling density and dataset. Yet, it remains important future work to assess performance under more drastic domain shifts (e.g., a change in anatomy between training-test sets). Recent studies have proposed adaptation of image priors to individual subjects to boost generalization to atypical anatomy \cite{narnhofer2019inverse,darestani2021accelerated}. During inference, adaptation methods optimize network parameters that form the prior so as to improve subject-specific reconstruction performance. While promising results have been reported for adapted priors, employing adaptation on diffusion priors with thousands of sampling steps substantially elevates computational burden \cite{gungor2022adaptive}. Although FDB improves efficiency by initiating sampling with a least-squares reconstruction, prior adaptation can be still challenging.

Several lines of limitations can be addressed to help boost the performance and utility of FDB in MRI reconstruction. A key area for improvement concerns learning strategies. Here, image priors were trained on MRI acquisitions pooled across multiple contrasts, but later used to reconstruct single-contrast data. Performance might be improved by employing contrast-specific modulation of feature maps for elevated specificity in the prior \cite{dalmaz2022one}, by training separate contrast-specific priors for reconstruction \cite{Dar2017}, or by performing joint reconstruction across multiple contrasts \cite{xiang2019rec,dar2020prior}. Moreover, here image priors were trained on datasets with fully-sampled acquisitions, which might limit utility in cases where acquisition of fully-sampled data is challenging \cite{yaman2020,liu2020rare}. The reliance of FDB on Nyquist-sampled acquisitions can be lowered by adopting self-supervised \cite{korkmaz2023selfsupervised,cui2022selfscore,aali2023solving,xiang2023ddm2}, cycle-consistent \cite{ozbey2022unsupervised} or low-rank assisted learning strategies \cite{peng2023oneshot}.

Other important areas for development concern the network architecture and frequency removal operator in FDB. Here, a UNet architecture common to diffusion modeling studies was used \cite{ho2020denoising}. Inverse problems in medical imaging have recently been reported to benefit from utilization of transformer backbones to capture long-range context \cite{korkmaz2022unsupervised,guo2022reconformer,feng2023mtrans}. Hybrid transformer-convolutional architectures might be utilized in FDB to achieve a favorable trade-off between spatial precision and contextual sensitivity \cite{dalmaz2021resvit}. Here, FDB assumed degradations based on spatial-frequency removal with peripheral-to-central ordering. This results in a variable-density sampling pattern for remaining frequency components at the start-point of the diffusion process, and FDB priors were later tested on variable-density undersampled acquisitions. The diffusion process in FDB can be adopted to implement other k-space sampling densities by modifying the scheduling of the k-space radius threshold over time steps accordingly. While our results suggest that FDB is reasonably reliable against domain shifts in sampling density, performance improvements can be sought by ensuring a match between training and test sets. It remains future work to assess the ideal architecture and frequency-removal scheduling for FDB-based MRI reconstruction.

\section{Conclusion}
In this study, we presented a novel constrained diffusion bridge, FDB, for accelerated MRI reconstruction. FDB's stochastic degradation operator performs random spatial-frequency removal to map directly between fully-sampled and undersampled data. For image reconstruction, FDB initiates sampling on the least-squares reconstruction of an undersampled acquisition, and performs reverse diffusion with a corrected sampling algorithm to gradually impute missing frequencies. Demonstrations on brain MRI indicate that FDB outperforms state-of-the-art reconstruction methods based on task-specific and task-agnostic priors.

\bibliographystyle{IEEEtran}
\bibliography{IEEEabrv,refs}

\end{document}